# 1/f spectral trend and frequency power law of lossy media

W. Chen

Simula Research Laboratory, P. O. Box. 134, 1325 Lysaker, Norway

(5 May 2003)

**The dissipation of acoustic wave propagation has long been found to obey an empirical power function of frequency, whose exponent parameter varies through different media. This note aims to unveil the inherent relationship between this dissipative frequency power law and 1/*f* spectral trend. Accordingly, the 1/f spectral trend can physically be interpreted via the media dissipation mechanism, so does the so-called infrared catastrophe of 1/*f* spectral trend[4]. On the other hand, the dissipative frequency power law has recently been modeled in time-space domain successfully via the fractional calculus and is also found to underlie the Levy distribution of media, while the 1/*f* spectral trend is known to have simple relationship with the fractal. As a result, it is straightforward to correlate 1/*f* spectral trend, fractal, Levy statistics, fractional calculus, and dissipative power law. All these mathematical methodologies simply reflect the essence of complex phenomena in different fashion. We also discuss some perplexing issues arising from this study.**

1. **Power law dissipation and 1/*f* spectral trend**

The dissipation of acoustic wave propagation through lossy media is generally described by $I = I_0 e^{-\alpha(f)t}$, where *I* represents the energy of an acoustic field variable (signal) such as velocity or pressure, *t* is the traveling time, and *f* frequency. The attenuation coefficient $\alpha(f)$ is experimentally found to obey a frequency power law function[1-3]

$$\alpha(f) = \alpha_0 f^y, \qquad y \in [0,2], \qquad (1)$$

where $\alpha_0$ and $y$ are empirical media-dependent parameters. For most solid and highly viscous materials, $y$ is close to 2, while for most of human tissues, $y$ ranges from 1 to 1.7. The underwater sediments and rock layers have $y \approx 1$, and for boundary layer loss of rigid tubes, $y=0.5$.

The power spectral (frequency) $P$ of a dissipative acoustic signal can be calculated by

$$P(f) = \int_0^{+\infty} I_0 e^{-\alpha_0 f^y t} dt = I_0 / \alpha_0 f^y. \qquad (2)$$

Note that the unit of power $P$ is the square of the signal unit rather than necessarily physical power (energy). It is seen from (2) that the frequency power law (1) of dissipative acoustic media underlies $1/f$ spectral trend, which means the power spectral of a signal is inversely proportional to the frequency $f$ according to a $1/f^\beta$ power law[4-6] ($\beta=y$, $P \propto 1/f^y$). Here $\beta$ is a nonnegative real number. $\beta=1$ is found in most phenomena, which causes the terminology "$1/f$ spectral trend". It is noted that a singularity develops at $\beta=1$. For $y=1$ dissipation, we also encounter the singularity issue in its corresponding fractional derivative model equation[7], and many biomaterials have the exponent $y$ around 1 (newly termed as y@1). It is also known[4] that the turbulent velocity of large Reynolds numbers follows the spectral trend $1/f^{5/3}$. Correspondingly, $y=5/3$ in terms of (1).

2. The infrared catastrophe

For the low frequency power $\int_0^1 P(f) df = I_0/\alpha_0 \int_0^1 1/f^y df$, $y \geq 1$ leads to infinite power (divergent), colorfully termed the infrared catastrophe[4]. The frequency power law of dissipation is found to take effect over a finite range of frequency, namely, $0 \prec f_{min} \leq f \leq f_{max} \prec \infty$, which echoes the self-similarity extends usually only over a finite range in real physical problems[8]. In terms of the power law dissipation, the $1/f$ power spectra do not hold for zero frequency component of a signal and avoids the infrared catastrophe.

On the other hand, the attenuation parameter expression $\alpha(f) = \alpha_1 + \alpha_0 f^y$, $y \in [0,2]$, also fits well with measured data of many dissipative acoustic signals[9], where $\alpha_1$ is also an empirical parameter. Accordingly, we have

$$\hat{P}(f) = \int_0^{+\infty} I_0 e^{-\alpha_1 - \alpha_0 f^y t} dt = \frac{I_0}{\alpha_1 + \alpha_0 f^y}. \tag{3}$$

Here the $1/f$ power spectral is revised as $1/(\alpha_1 + \alpha_0 f^\beta)$. It is obvious that (3) circumvents the infrared catastrophe. It is also interesting to connect the power law dissipation with the Wiener-Kinchin spectrum function in terms of page 218-242 of Mandelbrot [4]. We can accordingly assume $\alpha(f) = \alpha_0 f^y Q(f)$, which varies very slowly when $f$ close to 0.

## 3. Hurst exponent

The Hurst exponent is an essential measure of the smoothness of fractal time series. It is well known that the exponent parameter $\beta$ of $1/f$ signal has a simple relation with the Hurst exponent $H$

$$\beta = 2H+1, \tag{4}$$

where $H \in [0,1]$. Thus, we have

$$y = 2H+1. \tag{5}$$

However, the analysis shows that the relationship formulas are not correct or at best only hold for a limit range of the value of $H$. For example, $\beta = y = 2$ underlies a Gaussian process, and the corresponding $H=0.5$ indicates the Gaussian Brownian motion. The formula (4) is right here. But for $y=1$, the formula (5) produces $H=0$, corresponding to the white noise[6]. However, it is known that $y=1$ dissipation underlies the Cauchy statistic distribution[7] and, under certain condition, it is a deterministic process rather than the white noise. $y=1$ should correspond to $H=1$ for a deterministic process rather then $H=0$. In fact, the relationship formula (4) has been controversial when $H=0$ and $\beta=1$ for years. $\beta=0$, corresponding to $H=-1/2$ via (4), is often considered the white noise in terms of the $1/f$ spectral trend. Therefore, the relationship formula (4) and (5) do not hold.

### 4. Fractional calculus model, correlation function, and fractal dimension

The linear wave equation model of frequency-dependent dissipation, developed by Chen and Holm [7], is stated as

$$\Delta p = \frac{1}{c_0^2}\frac{\partial^2 p}{\partial t^2} + \frac{2\alpha_0}{c_0^{1-y}}\frac{\partial}{\partial t}(-\Delta)^{y/2} p, \qquad (6)$$

where $p$ is the pressure signal, and $(-\Delta)^{y/2}$ is the fractional Laplacian of the order $y$. The second right-hand term is to describe the dissipation of arbitrary frequency dependency, whose order is the value of $y$ as in the frequency power law (1). According to the equivalence of exponent parameters of the $1/f$ spectral trend and the power law dissipation, one can find the simple link between the fractional calculus model and the $1/f$ spectral trend.

By using both the fractional space/time derivatives, we have the fractional diffusion wave equation[10,11]

$$\frac{\partial^\mu p}{\partial t^\mu} = -\gamma(-\Delta)^{s/2} p$$
$$0 \leq s \leq 2,\ -1 \leq \mu \leq 2,\ 0 \leq y = s - \mu + 1 \leq 2, \qquad (7)$$

where $\gamma$ is the viscous constant, $s$ and $\mu$ are real number. The Fourier analysis[12] shows that (7) satisfies the power law (1). For $\mu=s=2$, equation (7) is the normal wave equation $\partial^2 u/\partial t^2 = \gamma\Delta u$ with $y=0$; for $\mu=1$, $s=2$, it is the normal diffusion equation $\partial u/\partial t = \gamma\Delta u$ with $y=2$; for non-integer $\mu$ and $s$, it is anomalous diffusion equation with $y=s-\mu+1$ (not for $s=\mu=1,2$) and essentially accounts for non-local and memory effects (entropy) on energy dissipations underlying a random walk (fractional Brownian motion)[13]. More precisely, the closer $y$ is to 1, the longer period dependency (stronger memory).

Let $s=2$, we have the fractional Signal equation

$$\frac{\partial^\mu p}{\partial t^\mu} = \gamma \Delta p . \tag{8}$$

In terms of (8), it is noted that $\mu=2H$, where $H$ is the Hurst exponent, seems reasonable. For $\mu=H=0$, the equilibrium equation $\Delta p = p$ underlies the frequency (time) independent white noise; for $\mu=1$ and $H=0.5$, the normal diffusion equation $\partial u/\partial t = \gamma \Delta u$ reflects the Gaussian process; for $\mu=2$ and $H=1$, the normal wave equation $\partial^2 u/\partial t^2 = \gamma \Delta u$ describes the non-dissipative deterministic wave propagation process. However, $0<\mu<1$, $0<H<0.5$ corresponds to $2<y<3$. This excessive dependency of dissipation on frequency is rarely, if ever, found in the real world and contradicts that $y=0$ underlies the frequency independent dissipation and white noise signal. This puzzle perplexes me deeply.

Equation (7) is also closely related to the autocorrelation function through

$$\langle x^2 \rangle = 2\int_0^t (t-\tau) corr(\tau) d\tau \propto t^{2\mu/s} , \tag{9}$$

where $corr(\tau)$ is the velocity autocorrelation function, and $\langle x^2 \rangle$ the position variance of random variable (mean square deviation). $0<\mu<1$ and $s=2$ indicates the sub-diffusion, while $\mu>1$, $s=2$ or $\mu=1$, $0<s<2$ means the super-diffusion process[10]. Is $H=\mu/s$?

Let $\mu=1$, we have the anomalous diffusion equation

$$\frac{\partial p}{\partial t} = -\gamma(-\Delta)^{s/2} p . \tag{10}$$

Observing (10), we find that $s \in [1,2]$ appears compatible with $H \in [1,0.5]$, $s \in [0,1)$ compares well with $H \in [0,0.5)$. There is the singular jump around $s=1$. Note that the exponent *y* in (1) equals *s* in terms of (10). For most media, *y* varies from 1 to 2, correspondingly, the majority of signals have *H* from 1 to 0.5.

It is well known that the Hurst exponent is related to the fractal dimension D by

$$D = d + 1 - H, \qquad (11)$$

where *d* is the topological dimension. The frequency dissipative power law (1) can be restated as

$$y = \frac{\ln \alpha(f)/\alpha_0}{\ln f}. \qquad (12)$$

(12) underlies a self-similarity with the fractal dimension *y*, called the dissipative dimension[7]. In terms of the connection between the 1/f spectral trend and multifractal[4], we assume that the dissipative dimension may collectively underlies the multifractal. The Gaussian dissipative dimension is 2 which echoes the second definition of the fractal dimension in page 38-39 of Mandelbrot [4]. Including the topological dimension, the fractal dimension is $D_y=d+2-y$, (How about $D_y=d+1-y/2$? No).

### 5. Levy stable distribution

In the context of kinetic physics, the frequency dissipative power law (1) also underlies the Levy *y*-stable statistic distribution[7], whose probability density function is the fundamental solution of the corresponding anomalous diffusion equation (10), where *s=y*. $y \in [0,2]$ coincides the fact that the Levy stable index ranges from 0 to 2. Accordingly, we have $\beta \in [0,2]$ of the spectral trend for dissipative acoustic signals in terms of the Levy stable distribution.

In terms of (10), the growth of the position variance can also be evaluated by the spatial Levy stable distribution function $S_y(x,t)$

$$\langle x^2 \rangle = \int x^2 S_y(x,t)dx \propto t^{2/y}. \qquad (13)$$

**6. Remarks**

Despite some perplexing issues, we have by far connect the $1/f$ power spectral, Hurst exponent, fractional calculus equation, fractal, and Levy stable process through the frequency power law of dissipative media. Let us summarize some definite results (1 dimension): 1) for a Gaussian process, $D=1.5$, $H=0.5$, $\beta=y=2$; 2) for a white noise process, $D=2$, $H=0$, $\beta=y=0$; 3); for a deterministic process, $D=1$, $H=1$, $\beta=y=1$. The $1/f$ power spectral is observed in a broad variety of physical, chemical, and biological phenomena. The present analysis shows that the frequency power law dissipation (1) is useful to describe all these $1/f$ power spectral signals.

The revelation of these underlying relationships has some potential applications. For instance, in the solution of inverse problem, we can first evaluate the exponent $\beta$ of the $1/f$ signal, and then have $y=\beta$. The signal noise distribution obeys the Levy $y$-stable distribution, the signal noise can be removed via $1/f^\beta$ filter, and the known parameter $y$ can directly be used to construct the corresponding linear or nonlinear dissipative wave equation model[7].

**Appendices:**

The power spectral can also be evaluated by the squared modulus of the Fourier transform, i.e., $FT\{I_0 e^{-\alpha_0 f^y t + ibft}\} = I_0(\alpha_0 f^y + bf + j\omega)/[(\alpha_0 f^y + j\omega)^2 + b^2 f^2]$, where $\omega$ is the angular frequency. It is noted that $y@1$, $H@0.75$, and $\beta@1$ (@ means $\approx$) have been observed in many cases. Is there a possible correspondence between $H \in [0.5,1]$ and $y \in [2,0]$, where $H=0.75,1$ respectively corresponds to $y=1,0$? In addition, may we assume $y=1/H$ for $1 \leq y \leq 2$ and $y=H/2$ for $0 \leq y < 1$ in terms of the fractal? Here we also dare to assume that the averaged dissipation of the high Reynolds number turbulence follows the Levy $y$-stable distribution.